\documentclass[aps,prl,reprint,groupedaddress,twocolumn,showkeys,amsmath,amssymb,longbibliography,floatfix]{revtex4-1}
\usepackage[english]{babel}
\usepackage{amsmath,amssymb,mathrsfs,bm,braket,color,graphicx,comment,amsfonts,mwe,tikz}
\usepackage{tabularx}
\usepackage[export]{adjustbox}
\usepackage[percent]{overpic}

\usepackage{contour,color,stackengine,titletoc,lipsum,tikz}
\usepackage[colorlinks,citecolor=blue,urlcolor=blue]{hyperref}
\usepackage[mathscr]{euscript}	
\usepackage[normalem]{ulem}

\newcommand{\e}{\text{e}}
\newcommand{\s}{\text{S}}
\newcommand{\ch}{\text{c}}
\newcommand{\n}{\text{n}}
\newcommand{\qp}{\text{qp}}

\begin{document}
\title{Identifying the $\nu=\frac{5}{2}$ topological order through charge transport measurements }

\author{Misha Yutushui}
\author{Ady Stern}
\author{David F. Mross}
\affiliation{Department of Condensed Matter Physics, Weizmann Institute of Science, Rehovot, 76100, Israel}
\date{\today}
\begin{abstract}
We propose an experiment to identify the topological order of the $\nu=\frac{5}{2}$ state through a measurement of the electric conductance of a mesoscopic device.
Our setup is based on interfacing $\nu=2, \ \frac{5}{2}$ and $3$ in the same device. 
Its conductance can unambiguously establish or rule out the particle-hole symmetric Pfaffian topological order, which is supported by recent thermal measurements. Additionally, it distinguishes between the Moore-Read and anti-Pfaffian topological orders, which are favored by numerical calculations.
\end{abstract}

\maketitle 

 {\bf Introduction}. Fractional quantum Hall states~\cite{Tsui_fqh_1982,Laughlin_fqh_1983,Haldane_fqh_1983,Halperin_fqh_1983} comprise many fascinating aspects of quantum many-body physics in an experimentally available platform. These features include topological order, which is reflected in fractional quantum numbers and statistics, and topologically protected edge states~\cite{Halperin_FQHE_2020}.
 The latter leads to robust experimental signatures such as a perfectly quantized electronic Hall conductance. A special group among the plethora of fractional quantum Hall states are those that exhibit highly-prized non-Abelian statistics~\cite{Wen_Non-Abelian_1991,nayak_non-abelian_2008,Moore_nonabelions_1991,Read_paired_2000}. The strongest numerical evidence of such a phase arises at the filling factor $\nu=\frac{5}{2}$~\cite{Morf_transition_1998,Rezayi_incompressible_2000,Peterson_Finite_Layer_Thickness_2008,Wojs_landau_level_2010,Storni_fractional_2010,Rezayi_breaking_2011,feiguin_density_2008,Feiguin_spin_2009}.
The nature of this phase remains hotly debated; the leading candidates based on numerical simulations are the celebrated Moore-Read (MR) Pfaffian~\cite{Moore_nonabelions_1991} and its particle-hole conjugate, known as anti-Pfaffian (APf)~\cite{Levin_particle_hole_2007,Lee_particle_hole_2007}.

The experimental distinction between different $\nu=\frac{5}{2}$ states is challenging due to the similarity of their edge structures: they feature the same charged modes and differ only in the number of Majorana modes. The parity of that number may be observed by interference measurements~\cite{Willett_5_2_2013,Stern_Probe_Non_Abelian_2006,Bonderson_Detecting_Non_Abelian_2006}. Quasiparticle tunneling has been proposed to identify the topological order via its scaling with temperature $T$ or gate voltage. However, experimentally determined exponents disagree with theoretical predictions, even for Abelian states~\cite{Chang_Chiral_LL_2003,Radu_Tunneling_2008,Lin_Measurements_2012}. These deviations may arise due to Coulomb interactions, details of the edge, or sample geometry~\cite{Tsiper_Formation_2001,Mandal_How_universal_2001,Mandal_relevance_2002,Wan_Universality_2005,Yang_Influence_2013}. Alternatively, thermal Hall conductance encodes a topological invariant that identifies the state. Ref.~\onlinecite{Banerjee_observation_2018} measured the two-terminal thermal conductance to be $\kappa \approx \frac{5}{2} $, in units of $\pi^2 k_B^2 T/3 h$. By contrast, the quantized values for MR and APf are $\frac{7}{2}$ and $\frac{3}{2}$. 
The measured value matches theoretical expectations for the PH-Pfaffian (PHPf) phase~\cite{Son_is_2015}, which is PH invariant\footnote{Generic states in this phase need not be symmetric~\cite{Zucker_stabilization_2016}, but map onto different state in the same phase.} but faces theoretical challenges~\cite{Antoni_Paired_2018,Stevan_model_Pf_2019}. It has never been observed numerically, and its trial states become critical upon projection into a single Landau level~\cite{Balram_parton_2018,Mishmash_numerical_2018,MY_Large_scale_2020,Rezayi_Energetics_2021}. Subsequent theoretical works argued that incomplete APf edge equilibration might also realize the measured value~\cite{Simon_equilibration_2018,Feldman_comment_2018,Simon_reply_comment_2018,Feldman_equilibration_2019,Simon_equilibration_2020,Asasi_equilibration_2020}, and shot-noise scaling was proposed to resolve this ambiguity \cite{Park_Noise_2020}.

Recent thermal noise measurements have strengthened the case for PHPf \cite{Dutta_novel_2021}. These experiments interfaced $\nu=\frac{5}{2}$ states with $\nu=3$ to realize the PH conjugate of the standard edge between $\nu=\frac{5}{2}$ and $\nu=2$. Observing  noise in both cases suggests that both `direct' and PH conjugate edges contain counterpropagating modes, which is the case only for PHPf.

\begin{figure}[b]
 \centering 
 \includegraphics[width=0.95\linewidth]{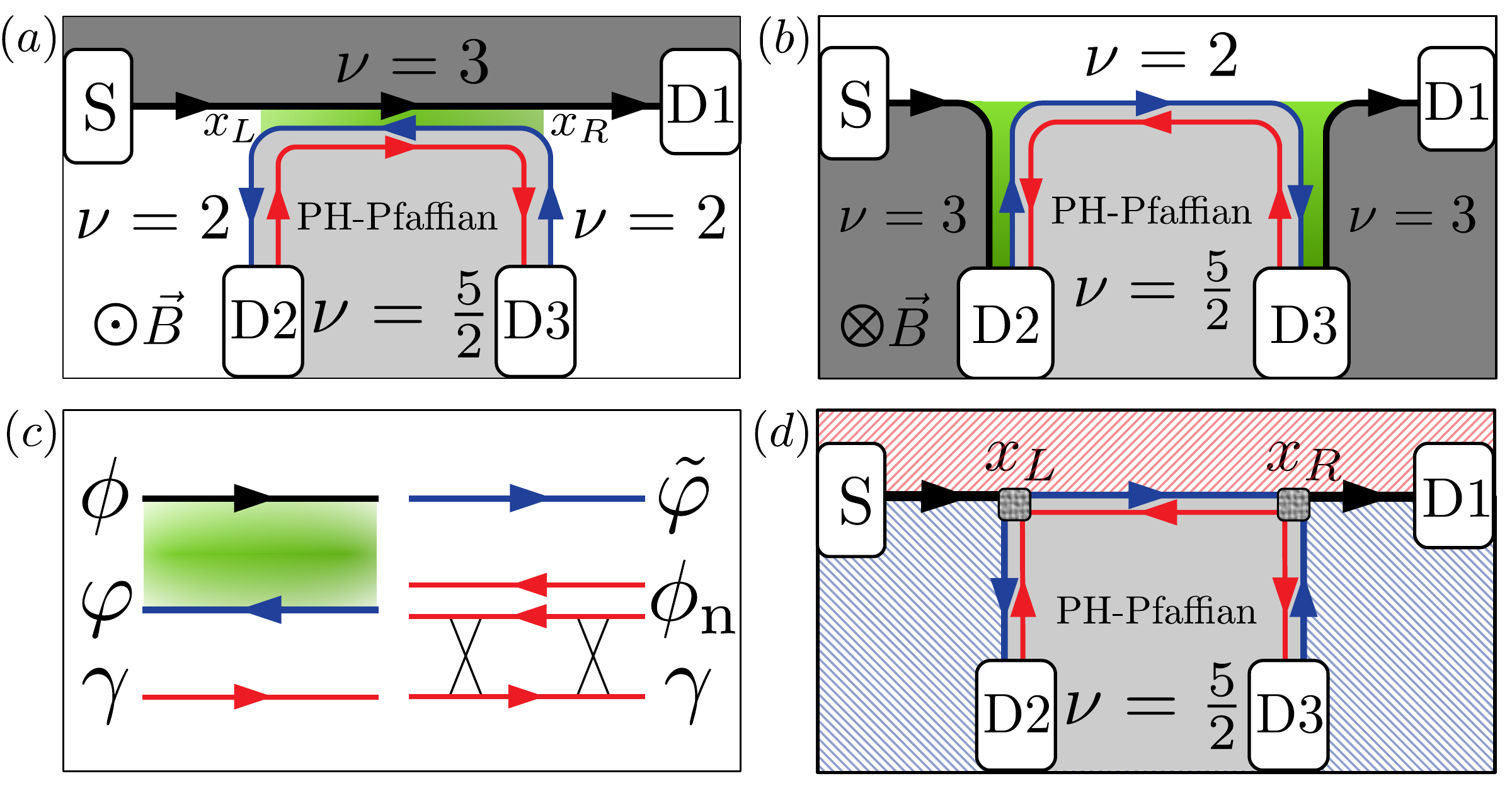}
 \caption{ Our setup exemplified for the PHPf topological order. The $\nu=\frac{5}{2}$ region is interfaced with $\nu=2$ and $\nu=3$. The positions of the integer states are interchanged between (a) and (b). Tunneling renormalizes the density-density interactions between electron ({\bf black}) and semion (\textcolor{blue}{blue}) modes to a fixed point with decoupled charge and neutral (\textcolor{red}{red}) modes (c). The neutral mode is partially compensated by the Majorana mode, leading to the same minimal edge structure in both cases (d). 
 \label{(a),(b),(c),(d)}}
 \label{fig.casesPH}
\end{figure}

Here, we show that devices that simultaneously contain interfaces of $\nu=\frac{5}{2}$ with $\nu=2$ and $\nu=3$ can provide critical information about the topological order at $\nu=\frac{5}{2}$.  Their electric conductances take different universal values depending on whether the edge with $\nu=2$ or with $\nu=3$ is chiral, i.e., carries only copropagating modes. The former includes the MR state, the latter the APf. Among the candidate topological orders, only PHPf features achiral edges with both $\nu=2,3$ and can therefore be identified uniquely.

In the setup of Fig.~\ref{fig.casesPH}, one electron mode emanates from the source S and enters the drain D1. Consequently, the conductance $G$ must lie between zero and one in units of $e^2/h$ (charge conservation implies that the conductance between S and D2 is $1-G$).  If it were solely determined by the bulk filling factors, the value $G=\frac{1}{2}$ would arise. We find that the conductance at $T=0$ can be either $G=\frac{1}{2}$ or $G=1$, and is dictated by the chirality of the neutral modes. Remarkably, it is insensitive to substantial deviations of tunneling exponents from their theoretically expected values. For the filling factors as in Fig.~\ref{fig.casesPH}(a), we predict complete transmission $G=1$ for MR but $G=\frac{1}{2}$ for PHPf and APf; in the PH-conjugate setup of Fig.~\ref{fig.casesPH}(b), we expect full transmission for APf and $G=\frac{1}{2}$ for both MR and PHPf. Thus, two charge measurements on the same device can distinguish between the three candidates. 

{\bf Quantum Hall states in the half-filled first excited Landau level}. The different electronic Hall conductances of the $\nu=\frac{5}{2}$ and $\nu=2$ states require a specific chiral charge mode at their interface. 
In addition, there are $n_M$ neutral chiral Majorana modes $\gamma_i$ that depend on the topological phase. The edge Lagrangian density reads $\mathcal{L}_\text{edge}=\mathcal{L}_0 + \delta \mathcal{L}$ with
 \begin{align}
 \mathcal{L}_0=\frac{\text{sgn}(v)}{2\pi}\partial_x\varphi [\partial_t-v\partial_x] \varphi+i\sum\limits_{l=1}^{n_M}\gamma_l [ \partial_t -v_l \partial_x]\gamma_l ~,\label{eq.action}
\end{align} 
where $s\propto e^{i\varphi}$ annihilates a charge-$e/2$ semion. For the MR state $n_M=1$, and the charge- and neutral-mode velocities $v,v_1$ have equal signs. Their signs are opposite for PHPf, which has $n_M=1$, and APf with $n_M=3$. The edge theory $\mathcal{L}_0$ admits interactions between charge and neutral modes, e.g., $\delta {\cal L}_1\propto\partial_x \varphi \gamma_i\gamma_j$ or $\delta {\cal L}_2\propto\partial_x \varphi \gamma_j \partial_x \gamma_j$. The scaling dimensions $[\gamma]=\frac{1}{2}$, $[s]=\frac{1}{4}$, and $[\varphi]=0$ imply that $\delta {\cal L}_1$ is marginal and $\delta {\cal L}_2$ is irrelevant in the RG sense. The former may change the edge conductance, while the latter does not affect macroscopic observables \footnote{Their implications for thermal equilibration are discussed in Ref.~\onlinecite{KKWM_Thermal_Equilibration_2020}}.

For our analysis, we also need the field theory of the particle-hole-conjugate edge. It is obtained by adding to ${\cal L}_\text{edge}$ a chiral mode of electrons $\psi^\dag \sim e^{i \phi}$ described by
\begin{align}\label{eq.electron}
{\cal L}_\text{e} =\frac{\text{sgn}(v_\e)}{4\pi}\partial_x\phi [\partial_{t}-v_\e\partial_x] \phi~,
\end{align}
with chirality opposite to the semion mode $\varphi$. The two edge theories ${\cal L}_\text{e},{\cal L}_\text{edge}$ couple via density-density interactions $\mathcal{L}_\text{int}=2h\partial_x\phi\partial_x\varphi$ and electron tunneling ${\cal L}_\text{tun}\propto i\sum_j \lambda_j\gamma_j\psi^\dagger s^2 $. The amplitudes $\lambda_j$ generically contain $x$-dependent phase factors, which cause destructive interference. Impurities result in randomly varying $\lambda_j(x)$, destroy this interference and facilitate tunneling between different edge modes \cite{Kane_Randomness_1994,Kane_Impurity_1995,Moore_Classification_1998,Moore_Critical_2002,Ferraro_Charge_2010,Protopopov_transport_2_3_2017}. Random tunneling is relevant when the scaling dimension of $\gamma_j\psi^\dagger s^2 $ is smaller than $3/2$. For $h=0$, we have $[\gamma_j\psi^\dagger s^2 ]_{h=0}=2$ and tunneling is irrelevant; it becomes relevant for  $h>h_\ch \equiv \frac{6-\sqrt{6}}{12}(v+v_\e)$ \cite{Levin_particle_hole_2007,Lee_particle_hole_2007}.

To analyze the latter case, we introduce new semions $\tilde{s} =\psi s^\dag$
and neutral fermions $\psi^\dag_\n=\psi^\dag s^2$.  These modes counterpropagate, with the chirality of $\tilde s$ opposite to that of $s$. The Gaussian terms in the composite-edge action comprise kinetic terms for $\tilde{s}^\dag\propto e^{i\tilde{\varphi}},\psi^{\dag}_\n\propto e^{i\phi_\n}$ and a coupling analogous to $\mathcal{L}_\text{int}$, which vanishes when $h=h_*\equiv \frac{2}{3}(v+v_\e)$\cite{Levin_particle_hole_2007,Lee_particle_hole_2007}. The tunneling for non-zero $n_M$ takes the form
\begin{align}\label{eq.tun}
 \mathcal{L}_\text{tun}= i\sum_{l=1}^{n_M}\lambda_l(x)\gamma_l\psi_\n + \lambda^*_l(x)\gamma_l\psi^\dag_\n~.
\end{align} 
When $\psi_\n$, which comprises two Majoranas, and $\gamma_l$ have opposite chirality, pairs of counterpropagating Majoranas `compensate' each other: They become localized at a scale $\xi$ and do not affect the physics at longer wavelengths~\cite{Giamarchi_Anderson_localization_1988}. The remaining $|2- n_M|$ (opposite signs of $v,v_1$) or $2+ n_M$ (equal signs) Majoranas enjoy topological protection. They may interact with the charge mode $\tilde \varphi$ through terms such as $\delta{\cal L}_{1,2}$. However, such couplings become irrelevant at a random edge~\cite{Levin_particle_hole_2007,Lee_particle_hole_2007}. Consequently, charge and neutral sectors decouple in all cases pertinent here.

{\bf PH-Pfaffian}. This topological order is unique in having the same minimal edge structures when interfaced with $\nu=2$ and $\nu=3$, up to a global chirality reversal. This property follows from Eq.~\eqref{eq.tun}, which localizes a pair of counterpropagating Majoranas and leaves behind an unpaired one propagating oppositely to $\tilde s$, see Figs.~\ref{fig.casesPH}(c,d).

In Fig.~\ref{fig.casesPH}(a), the PHPf island is immersed in a $\nu=2$ region adjacent to a $\nu=3$ region. We consider the case where D1,D2,D3 are grounded, while the source is at a potential $V$. The conductance measurement is then a scattering experiment, with an electron emanating from the source as the incoming state. Outgoing states must carry the electron charge either as an electron going to D1 or two semions flowing to D2. In the latter case, a Majorana fermion must enter D3 to conserve fermion parity. We use the currents at S,D1,D2,D3 as
boundary conditions for the segment between $x_{R/L}$ to calculate the conductance. The current emanating from S propagates unimpeded to $x_L$ and is carried by electrons. As such, it is $I_\e^S=\partial_t\phi(x_L)/2\pi$. Similarly, the current that enters D1 is $I_\e^{D1}=\partial_t\phi(x_R)/2\pi$. The currents from D3 and into D2 are given by $I_\qp^{D2,D3}=-\partial_t\varphi(x_{L,R})/2\pi$. 

In the middle region, the charge mode $\tilde{s}$ and neutral mode $\psi_\n $ decouple. The corresponding charge current is $I_\ch\equiv \partial_t\tilde{\varphi}/2\pi $, and the neutral current is $I_\n\equiv -\partial_t\phi_\n/2\pi$. The boundary conditions are thus
\begin{equation}\label{eq.contin}
\begin{split}
 &I_\ch(x_L)=I_\e^\s-I_\qp^\text{D2},\qquad\ \
 I_\n(x_L)=I_\e^\s-2I_\qp^\text{D2},\\
 &I_\ch(x_R)=I_\e^\text{D1}-I_\qp^\text{D3},\qquad
 I_\n(x_R)=I_\e^\text{D1}-2I_\qp^\text{D3}.
\end{split}
\end{equation}
Charge conservation implies $I_\ch(x_R)=I_\ch(x_L)$, but the neutral current is sensitive to interactions at the interface. In the absence of the tunneling term ${\cal L}_\text{tun}$, $I_\n(x_L) = I_\n(x_R)$, thus, any current emanating from S flows into D1. By contrast, when tunneling localizes one of the Majorana modes, $I_\n(x_L)$ vanishes exponentially with $|x_L-x_R|/\xi$. A neutral fermion at $x_R$ has an equal probablitity for arriving at $x_L$ as a particle or hole \cite{Chung_Majorana_backscattering_2011}. Consequently, the source current splits equally between D1 and D2 such that $G=\frac{1}{2}$~\cite{Maslov_Landauer_conductance_1995,Safi_Transport_1995,Ponomarenko_Renormalization_1995}.

 \begin{figure}[t!]
 \centering
 \includegraphics[width=0.95\linewidth]{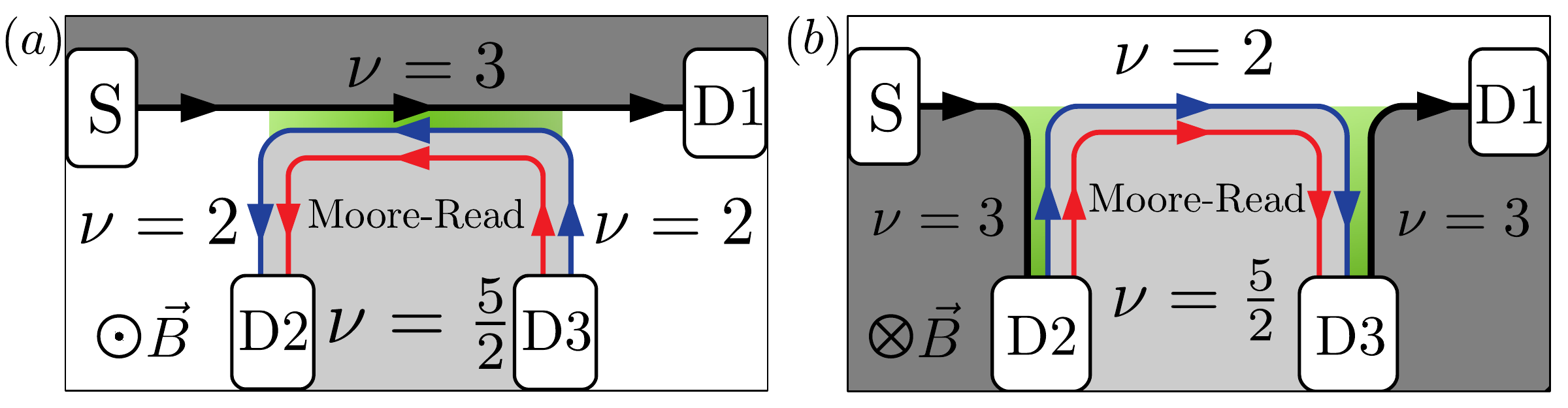}
 \caption{The same setup as in Fig.~\ref{fig.casesPH} for the MR topological order. (a) The tunneling between electron mode and MR edge is irrelevant, leading to total transmission, i.e., $G=1$ at $T=0$ (b) Tunneling drives the edge connected to D2,D3 to a fixed point with decoupled charge and neutral modes and $G=\frac{1}{2}$.}
 \label{fig.casesMR}
\end{figure}

One may worry that local processes that couple the neutral and charge modes lead to $I_\n(x_L)\ne 0$ and modify the conductance. The most dangerous coupling is an interaction between charge and neutral modes of the form $\partial_x \tilde \varphi e^{2i\phi_\n}+H.c.$ It is, however, irrelevant and does not affect the conductance at low voltages. 

The relation between the minimal edge structures of $\nu=2,3$ interfaced with PHPf implies that the same result holds for the PH-conjugate setup of Fig.~\ref{fig.casesPH}(b). It is instructive to analyze this setup starting from the limit where the original edge modes are decoupled. Then, all the current that emanates from S enters D2 and $G=0$. When the edges couple through density-density interactions only, the eigenmodes between $x_{L,R}$ and D2,D3 are $h$-dependent linear combinations of the counterpropagating electron and semion modes. There is a flow of current from S to D1, with an $h$--dependent non-universal conductance \cite{Supplemental_Material}. For $h=h_*$, the eigenmodes are the charge and neutral modes, and $G=1/2$. Strong tunneling localizes one of two Majorana fermions comprising $\psi_n$. It thereby prohibits marginal couplings between charge and neutral sectors, enforcing a universal conductance $G=1/2$.

{\bf Moore-Read and anti-Pfaffian}. We turn to a MR state in the configuration of Fig.~\ref{fig.casesMR}(a). Its interface with $\nu=3$ is described by Eqs.~(\ref{eq.action},\ref{eq.electron},\ref{eq.tun}). Crucially, the neutral fermions $\psi_\n$ and $\gamma$ copropagate and cannot become localized by Eq.~\eqref{eq.tun}. Instead, this term renormalizes $h$ to $h_*$, thus decoupling charge and neutral modes \cite{Levin_particle_hole_2007,Lee_particle_hole_2007}. We compute the conductance using Eq.~\eqref{eq.contin}. Without tunneling, $I_\n(x_L)=I_\n(x_R)$ and $G=1$. When tunneling is allowed, a neutral fermion at $x_R$ has a non-zero bare amplitude for arriving at $x_L$ as a hole. At voltages beyond a scale $\propto |x_L-x_R|^{-1}$ this leads to a non-universal differential conductance. At low voltages, the bare amplitude is renormalized towards a fixed-point value.

Low-energy modes experience the interface region as effectively point-like. To identify fixed points and find the nearby flow, we replace the interface with boundary conditions for the lead variables at $x_{R,L}$. The trivial fixed point without tunneling is encoded by $\psi_\n(x_L) = \psi_\n(x_R)$ and exhibits $G=1$~\cite{Maslov_Landauer_conductance_1995,Safi_Transport_1995,Ponomarenko_Renormalization_1995,Protopopov_transport_2_3_2017}. The opposite limit corresponds to $\psi_\n(x_L) = \psi_\n^\dagger(x_R)$. Here, Eq.~\eqref{eq.contin} with $I_\n(x_L)=-I_\n(x_R)$ yields $G=\frac{1}{3}$ (see also \cite{Supplemental_Material}). Analogous fixed points were found in the context of $\nu=2/3$ in Ref.~\onlinecite{Protopopov_transport_2_3_2017}. Their stability is determined by the most prominent local perturbation, the tunneling term ${\mathcal L}_\text{tun}$, now acting at one point $x_L\approx x_R$. At the trivial fixed point, the neutral fermion is expressible in terms of incoming modes via $\psi_\n=s^2(x_R)\psi^\dagger(x_L)$. As such, its scaling dimension is $[\psi_\n]=3/2$, while $[\gamma]=1/2$ as before. Consequently, this perturbation is irrelevant, and the $G=1$ fixed point is attractive. Its stability also follows directly from the electron scaling dimension at an isolated fractional edge, which is $[s^2\gamma]=3/2$ according to Eq.~\eqref{eq.action}. However, stability only requires $[s^2\gamma]>1/2$, such that our results are robust to significant deviations from the theoretical value  (see also \cite{Supplemental_Material}). At the non-trivial fixed point, the neutral fermion satisfies $\psi_\n^{\dagger 3}=s^{2}(x_R)\psi^\dagger(x_L)$ and thus $[\psi_\n]=1/6$. Here, the perturbation is relevant, and the fixed point is repulsive~\cite{Supplemental_Material}. Consequently, the conductance is generically determined by the trivial fixed point with $G=1$.

The analysis of the MR state in the geometry of Fig.~\ref{fig.casesMR}(b) mirrors that of the PHPf. Without inter-edge coupling $G=0$, while density-density interactions without tunneling lead to a non-universal $h$-dependent conductance. Strong tunneling results in $h=h_*$~\cite{Lee_particle_hole_2007,Levin_particle_hole_2007}, for which $G=1/2$. Local perturbations near $x_L$ are irrelevant and do not modify the conductance at low voltage.

The properties of the APf phase follow from a global PH-conjugation $\nu=3\leftrightarrow \nu=2$. Consequently, $G=\frac{1}{2}$ in the geometry of Fig.~\ref{fig.casesMR}(a) and $G=1$ for Fig.~\ref{fig.casesMR}(b).

\begin{figure}[t!]
 \centering
 \includegraphics[width=0.95\linewidth]{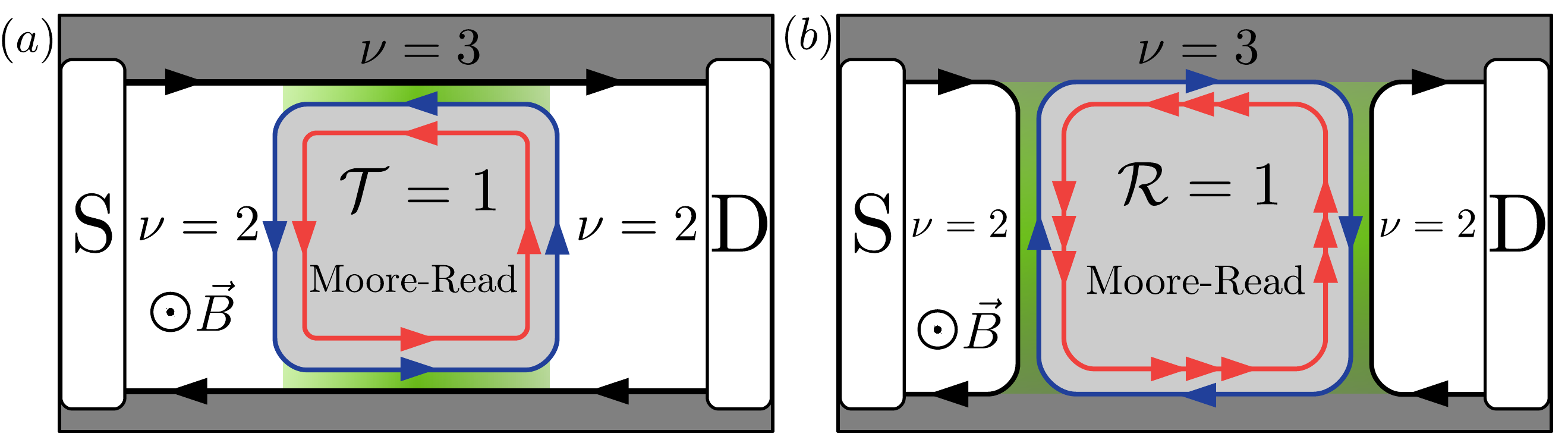} 
 \caption{The two-terminal setup with the $\nu=\frac{5}{2}$ island in the MR phase, embedded in a $\nu=2,3$ environment. (a) The fully transmitting fixed point is stable in the limit $L_y\gg L_x$. (b) The fully reflecting  fixed point is unstable in the limit $L_y\gtrsim \xi$.}
 \label{fig.Island}
\end{figure}

{\bf Island geometry.}
As an alternative to the geometries of Figs.~\ref{fig.casesPH}, \ref{fig.casesMR} one may consider a $\nu=\frac{5}{2}$ island of size $L_x \times L_y$ `floating' on a $\nu=2,3$ background, see Fig.~\ref{fig.Island} and Refs. \onlinecite{Rosenow_signatures_2010,Protopopov_transport_2_3_2017}. Here, the source and the drain are connected only to integer modes. For sufficiently low energies, the fractional island acts as a local scatterer with transmission ${\cal T}(\omega)$ and reflection ${\cal R}(\omega)$ probabilities that depend on the incident electron energy. When the island is large compared to microscopic length scales, the bare probability amplitudes are renormalized. On general grounds, we expect the existence of trivial fixed points with ${\cal T}=1$ and ${\cal R}=1$, see Fig.~\ref{fig.Island}. Since the leads contain only integer modes, they do not cause renormalization, and the island’s dimensions cut off the RG-flow.

When examining the stability of both fixed points, we find that for the MR and APf states, one is stable, while the other is unstable. By contrast, both are unstable for the PHPf, necessitating the existence of an additional, non-trivial fixed point. The qualitative behavior near the trivial fixed points is straightforward: For weak tunneling along the horizontal path, the island exhibits discrete energy levels with splitting $\Delta E \propto \frac{1}{L_x + L_y}$. Transmission past the island is nearly perfect unless the energy of the scattering electron matches one of these levels. The conductance exhibits sharp dips at those energies, which broaden as the integer modes hybridize  with the island. In the opposite limit, there is near-perfect reflection, and the conductance is low apart from sharp resonance peaks. For a particular topological phase of the island, the renormalization of tunneling determines which limit is realized. 

Consider first a MR island. Near the ${\cal T}=1$ fixed point, Fig.~\ref{fig.Island}(a), $\mathcal{L}_\text{tun}$ contains only copropagating modes. The horizontal edge is characterized by decoupled charge and neutral modes. This segment is governed by a stable fixed point where no charge transfers between the electron and semion modes. For $L_x \ll L_y$, renormalization due to the vertical edges is effective, and the resonances \textit{sharpen} at low incident energies. The transmission is thus close to unity at generic energies. When $L_y \ll L_x$, renormalization is inefficient, and deviations from complete transmission are significant. The ${\cal T}=1$ fixed point is thus (marginally) stable.
 
In contrast, at the ${\cal R}=1$ fixed point, Fig.~\ref{fig.Island}(b), the island decouples from the $\nu=2$ region. Tunneling described by Eq.~\eqref{eq.tun} is strongly relevant, and two pairs of Majorana modes localize at the scale $\xi$. Thus, the decoupling becomes unstable if $L_y\gtrsim\xi$ and the transmission resonances broaden at low incident energies. For APf, the two cases are reversed, and ${\cal R}=1$ is marginally stable.

For PHPf, we again start near the fixed point ${\cal T}=1$. Similar to the MR case, tunneling is strongly relevant. It leads to the localization of a pair of Majorana modes at the length $\xi \ll L_x$, which destroys the fully transmitting fixed point. The same reasoning also holds near ${\cal R}=1$, i.e., both trivial fixed points are unstable. Therefore, when $L_x,L_y \gg\xi$, the transport must be governed by an intermediate, stable fixed point. Since the details of the island drop out in this limit, we expect this fixed point to exhibit the universal conductance $G=1/2$ and no resonances. 

{\bf Non-zero temperatures.} Up to this point, our analysis assumed $T=0$, where all transport is fully coherent. At low temperatures, the results still hold with power-law corrections governed by the leading irrelevant operators at the respective fixed points.

When $T$ is large enough that the thermal length $L_T\propto 1/T$ is much shorter than the length $|x_R-x_L|$ of the interface (while still much lower than the bulk gap for delocalized excitations), the system behaves like a network of classical resistors. When dephasing is strong enough to establish a local chemical potential at $x_L$ [see Fig.~\ref{fig.casesPH}], the current arriving from the source splits equally between the arms leading to D1,D2, resulting in $G=1/2$ for all candidate states, just as the low-temperature result for the PHPf (see also Ref.~\onlinecite{Lai_Distinguishing_2013}, which studied a related geometry). In principle, the temperature dependence of $G$ distinguishes between the coherent and incoherent origins of $G=1/2$. While in the former case, deviations from $1/2$ increase with temperature, in the latter, they decrease. Alternatively, coherence may be probed by incorporating an interference loop for the integer states.

 {\bf Discussion}. We have proposed two experimental setups where coherent charge-transport measurements can distinguish between three classes of $\nu=\frac{5}{2}$ states. (i) Those whose interfaces with $\nu=2$ are chiral, e.g., MR. (ii) States that exhibit chiral interface with $\nu=3$, e.g., APf. (iii) The PHPf whose interfaces with both $\nu=2,3$ are non-chiral. Within each setup, two charge measurements can uniquely identify the class of the state, see Tab.~\ref{tab.tab}. Both setups require a specific hierarchy of length scales: The localization length $\xi$ must be the shortest scale to guarantee that each edge state is reduced to its topologically required minimum. The thermal length must be the longest scale to ensure coherent transport. Additionally, the island geometry requires measurements for both limits of the aspect ratios $L_x \gg L_y$ and $L_y \gg L_x$, or for two configurations related to one another by $\nu=2 \leftrightarrow \nu=3$.
 
Finally, the same information about the topological order may be obtained from a setup where the $\nu=2$ state is substituted by $\nu=0$. The additional integer edge modes add $2$ units to the conductance in the setup referred to in the second and fourth columns of Table (\ref{tab.tab}). In the same way, one can replace $\nu=3$ by a larger integer.

\begin{table}[t!]
 \centering
 \caption{ The conductance between source S and drain D1 for two different sets of filling factors, with non-univ. standing for `non-universal' conductance bounded between 0 and 1~\cite{Rosenow_signatures_2010}. The last column corresponds to the setup of Fig.~\ref{fig.Island} with interchanged $\nu=2$ and $\nu=3$ states. }
 \begin{tabular}{ccccc}
 \hline \hline 
 & Fig.~\ref{fig.casesMR}(a) & Fig.~\ref{fig.casesMR}(b) & Fig.~\ref{fig.Island} & Fig.~\ref{fig.Island}\;$ (\;2\leftrightarrow 3\;)$ \\\hline
 Moore-Read & 1 & 1/2 & 1 & non-univ. \\
 PH-Pfaffian & 1/2 & 1/2 & 1/2 & 1/2\\
 anti-Pfaffian & 1/2 & 1 & non-univ. & 1\\\hline \hline 
 \end{tabular}
 \label{tab.tab}
\end{table}

\begin{acknowledgments}
{\bf Acknowledgments: }
It is a pleasure to thank  Moty Heiblum and Bivas Dutta for illuminating discussions on this topic. This work was partially supported through CRC/Transregio 183, by the ERC (LEGOTOP), the ISF (1866/17) and within the ISF-Quantum program.
\end{acknowledgments}

\bibliography{phpfbib}

\end{document}